# Evolution of superconductivity in $PrFe_{1-x}Co_xAsO$ (x = 0.0 – 1.0)

Poonam Rani[1], Anand Pal[1,2] and V. P. S. Awana[1,*]

[1]Quantum Phenomena and Application Division, National Physical Laboratory (CSIR)
Dr. K. S. Krishnan Road, New Delhi-110012, India
[2]Department of Physics, Indian Institute of Science, Bangalore- 560012

## ABSTRACT

We report the synthesis and physical property characterization of $PrFe_{1-x}Co_xAsO$ (x = 0.0 – 1.0). The studied samples are synthesized by solid state reaction route via vacuum encapsulation method. The pristine compound PrFeAsO does not show superconductivity, but rather exhibits a metallic step like transition due to spin density wave (SDW) ordering of Fe moments ($Fe^{SDW}$) below 150 K, followed by another upward step due to anomalous ordering of Pr moments ($Pr^{TN}$) at 12 K. Both the $Fe^{SDW}$ and $Pr^{TN}$ temperatures decrease monotonically with Co substitution at Fe site. Superconductivity appears in a narrow range of x from 0.07 to 0.25 with maximum $T_c$ at 11.12 K for x = 0.15. Samples, with x ≥ 0.25 exhibit metallic behavior right from 300 K down to 2 K, without any $Fe^{SDW}$ or $Pr^{TN}$ steps in resistivity. In fact, though $Fe^{SDW}$ decreases monotonically, the $Pr^{TN}$ is disappeared even with x = 0.02. The magneto transport measurements below 14 Tesla on superconducting polycrystalline Co doped PrFeAsO lead to extrapolated values of the upper critical fields [$H_{c2}(0)$] of up to 60 Tesla.

Key Words:

A. Fe based Pnictides
B. Solid State Synthesis
D. Superconductivity
E. Magneto-transport

## Introduction

The discovery of iron-based layered superconductors has attracted tremendous interest of the solid state physics researchers [1]. The most intriguing part is that these compounds contain familiar ferromagnetic Fe and yet exhibit superconductivity of up to above 55 K, which is second only to high $T_c$ cuprates [2-7]. The un-doped non superconducting ground state of these compounds i.e., REFeAsO show an anomaly in resistivity measurements below 150 K due to coupled crystallographic change and the spin density wave (SDW) like magnetic transition of Fe spins [2-9]. Superconductivity is achieved along with simultaneous suppression of crystallographic and magnetic phase transitions, by carrier doping via O site F substitution or vacancies in RE-O layer [2-7]. Later, it was observed that superconductivity can be induced in these compounds by direct injection of electrons in the conducting FeAs, viz substitution of Co at Fe site [10-15]. As far as doping mechanism is concerned, some similarities have been seen in Fe and Cu based superconductors. In both class of compounds the superconductivity is achieved by inducing electron or hole carriers from adjacent redox layers to the two-dimensional conduction layers, i.e., FeAs in oxy-pnictides and $CuO_2$ in HTSc Cuprates. Though, in case of HTSc cuprates the direct doping of carriers by Cu site substitutions was not possible, but the same is possible in case of Fe based pnictides with in same structure with full Fe site Co or Ni substitutions [15-18].

In view of our earlier work on evolution of superconductivity by Fe site Co doping in $SmFe_{1-x}Co_xAsO$ [11] and the existence of only scant reports [19] on $PrFe_{1-x}Co_xAsO$ in low doping range, prompted us to study the doping of Co into the superconducting-active FeAs layer of non superconducting PrFeAsO and the evolution of superconductivity. Superconductivity has been observed in narrow window of Co doping i.e. $0.07 \geq x \leq 0.25$ with the maximum $T_c$ 11.12 K at $x = 0.15$. We present the complete phase diagram of $PrFe_{1-x}Co_xAsO$, with x ranging from 0.0 to 1.0

**Experimental**

The studied polycrystalline $PrFe_{1-x}Co_xAsO$ (x = 0.0, 0.02, 0.05, 0.07, 0.10, 0.15, 0.20, 0.25, 0.30, 0.50, 0.75 and 1.0) samples are synthesized through single step solid-state reaction route via vacuum encapsulation technique. High purity (~99.9%) powders of Pr, As, $Fe_2O_3$, Fe, and $Co_3O_4$ in their stoichiometric ratios are weighed, mixed and ground thoroughly using mortar and pestle in presence of high purity Ar atmosphere in glove box. The Humidity and Oxygen content in the glove box are maintained at less than 1 ppm. The mixed powders are palletized and vacuum-sealed ($10^{-4}$ Torr) in a quartz tube. These sealed samples are placed in box furnace under the heat treatment, 550ºC for 12 h, 850ºC for 12 h and then at 1150ºC for 33 h in continuum with slow heating rate. Finally furnace is finally cooled down to room temperature. The crystal structure and phase are analyzed by the powder X-ray diffraction patterns at room temperature using Rigaku X-ray diffractometer with $CuK_\alpha$ radiation. The resistivity measurements under magnetic field are carried out by a conventional four-probe method using Quantum Design Physical Property Measurement System (PPMS) with fields up to 14 T.

**Results and Discussion**

Fig. 1 (a) shows the Rietveld fitted X-ray diffraction (XRD) patterns of the representative samples of $PrFe_{1-x}Co_xAsO$ (x = 0.0 - 1.0) series. It is clear from Figure 1 that all the samples are single phase and are crystallized in the tetragonal phase with space group *P4/nmm* in analogy of ZrCuSiAs type structure. Impurity lines are not seen in the detectable range within x-ray resolution, except a few very minor peaks close to the background. *FULLPROF SUITE* program is used to refine the room temperature X-ray diffraction patterns of the studied samples. The Rietveld analysis is carried out in the space group P4/*nmm*. The Wyckoff positions for Pr and As are taken to be located at 2*c* (1/4, 1/4, *z*), the O at 2*a* (3/4, 1/4, 0) and the Fe/Co are shared at site 2*b* (3/4, 1/4, 1/2). The details of refined cell parameters along with the quality of fitting parameter are listed in the Table1. The variation of lattice parameters and unit cell volume as function of Co-doping (*x*) are plotted in Fig. 1 (b). It is observed that the volume and the c -lattice parameter decrease with increase in Co substitution at Fe site. The decrement in both parameters infers that the Co is successfully substituted at Fe-site, which is in agreement with other reports on $REFe_{1-x}Co_xAsO$ [10-19].

Figure 2 shows the temperature dependence of normalized resistivity ($\rho/\rho_{300}$) for Co doped $PrFe_{1-x}Co_xAsO$ samples with x from *x* = 0.0 to 1.0. The temperature dependent resistivity $\rho(T)$ curve of un-doped PrFeAsO sample exhibits slightly semiconducting behavior from room temperature down to around 150 K and below this temperature the resistivity becomes metallic. This anomaly in the resistivity is the result of the coupled structural phase transition from the tetragonal *P*4/*nmm* to the orthorhombic *Cmma* space group at around T~150 K, and the commensurate anti-ferromagnetic (AF) magnetic ordering of the Fe spins at a slightly lower temperature of ~140K, also called 'spin density wave' (SDW) like transition. [2-9]. An upward resistivity step at around 12 K is also observed, which could be clearly seen as a kink in d$\rho$/dT plot, the same is not shown here as was recently reported by some of us elsewhere for the same sample, please see ref. 20. The 12 K resistivity step is seen earlier in resistivity data by others as well and was assigned to the unusual antiferromagnetic ordering of Pr moments ($Pr^{TN}$) due to $Pr^{4f}$ and $Fe^{3d}$ interaction [21, 22]. The situation is similar to the case of infamous non superconducting HTSc cuprate $PrBa_2Cu_3O_7$, where $Pr^{4f}$ and $Cu^{3d}$ interaction gave rise to anomalous $Pr^{TN}$ of around 17 K [23]. Interestingly the $Pr^{TN}$ is disappeared with 2% doping of Co at Fe site in PrFeAsO. The SDW transition temperature also decrease drastically with Co-substitution and the same becomes 54.66K at 2% Co doping and is not seen down to 2 K for 5% Co doped sample. The variation of $Fe^{SDW}$ with x is shown in inset of Fig.2. The 7% Co – doped sample shows onset of superconducting transition temperature at around 7 K, but does not show zero resistivity down to 2 K. The resistivity curves of $PrFe_{0.90}Co_{0.10}AsO$ and $PrFe_{0.85}Co_{0.15}AsO$ show metallic behavior from room temperature down to superconducting transition onset of around 13.3 K. Interestingly the resistivity of 15% Co-doped sample is more metallic in comparison to 10% Co-doped one and the $T_c$ ($\rho$=0) is seen at 11.12 K and 10.16 K respectively. The 20% and 25% Co-doped sample though exhibited onset of the superconductivity at 8.6K and 4K but did not show zero resistivity down to 2 K. On further doping (x > 0.25), the metallic behavior of the compound becomes more prominent, but the superconductivity is disappeared. This is due to the over doping of carriers in the conducting plane.

Inset of Fig. 2 shows the superconducting transition temperature dependence on Co concentration (x), demonstrating a domelike curve with highest $T_c$ at 11.16 K for the optimal doping of x = 0.15. The superconducting transition temperature dependence on Co-doping(x) is a parabolic curve with highest $T_c$ at 11.16K for x = 0.10. Above this doping the $T_c$ starts to decrease. It is clear that, PrFe$_{1-x}$Co$_x$AsO shows the narrow superconducting window. Samples with x > 0.25 show linear metallic behavior right from 300 K down to 2 K without any Fe$^{SDW}$ or Pr$^{TN}$ steps in resistivity and are not superconducting. In general, the appearance and disappearance of T$_c$ with x in PrFe$_{1-x}$Co$_x$AsO is similar to that as reported previously for other REFe$_{1-x}$Co$_x$AsO compounds [10-15].

Figures 3 (a) and (b) show the temperature dependence of normalized resistivity $\rho/\rho_{14}$ of PrFe$_{0.9}$Co$_{0.1}$AsO and PrFe$_{0.85}$Co$_{0.15}$AsO samples under applied magnetic fields up to 10 Tesla. Field is applied perpendicular to the direction of current flow in the samples. It is clear from the data that the resistive transition shifts towards the lower temperature with increase in magnetic field. We can see that the onset part is almost unaffected by the applied magnetic field but the offset is decreased substantially. The onset transition temperature (T$_c^{onset}$) for 10% Co doped sample of 13.4 K (0 Tesla) is shifted to 12.3 K (10 Tesla) and for 15% Co doped PrFeAsO sample the same shifts from 13.5 K (0 Tesla) to 12.4 K (10 Tesla). On the other hand noticeable shift has been observed on offset temperature i.e., from 10.16 K (0 Tesla) to around 3.0 K (10 Tesla) for 10% Co doped sample and 11.16 K (0 Tesla) to 4.5 K (10 Tesla) for 15% Co doped sample. The rate of decrease of offset transition temperature with applied magnetic field of the Co-doped PrFe$_{0.9}$Co$_{0.1}$AsO and PrFe$_{0.85}$Co$_{0.15}$AsO superconducting samples is around 0.72 K/Tesla (dT$_c$/dH = 0.61 K/Tesla) and 0.66 K/Tesla (dT$_c$/dH = 0.74 K/Tesla) respectively, which are far less in compare to other high T$_c$ superconductor like, YBCO (dT$_c$/dH = 4 K/Tesla) [24] and MgB$_2$ (dT$_c$/dH = 2 K/Tesla) [25]. The lower value of dT$_c$/dH indicates towards the robustness of superconductivity in these compounds and higher value of the upper critical field (H$_{c2}$) values.

The upper critical field [H$_{c2}$(T)] values at zero temperature are calculated by the extrapolation method using Ginzburge-Landau (GL) theory. The H$_{c2}$(T) is determined using different criterion of H$_{c2}$(T) = H at which $\rho$ = 90%, 50% and 10% of $\rho_N$, where $\rho_N$ is the normal state resistivity. The Ginzburge-Landau equation is:

$$= \quad (0)*$$

where, t = T/T$_c$ is the reduced temperature and H$_{c2}$(0) is the upper critical field at temperature zero. The variation of H$_{c2}$(T) with temperature for 10% and 15% Co doped sample is shown in Fig. 4(a) and (b) respectively. The H$_{c2}$(0) of 10% Co doped PrFeAsO sample reaches above 60 Tesla with 90% criteria and the same is around 54 Tesla for 15% Co doped PrFeAsO.

The temperature derivative of resistivity (d$\rho$/dT) versus temperature (T) plots for the superconducting samples PrFe$_{0.9}$Co$_{0.1}$AsO and PrFe$_{0.85}$Co$_{0.15}$AsO at various applied magnetic field are shown in Fig. 5 (a) and 5 (b). A narrow intense peak is observed at T$_c$ at zero applied fields indicating good percolation path of superconducting grains. The resistivity peak is broadened under applied fields and broadening of the d$\rho$/dT peak increases with applied magnetic field. The broadening of resistivity in superconductors under applied magnetic field has been reported to be caused by the creep of vortices, which are thermally activated for both single crystal and polycrystalline bulk sample [26, 27]. Therefore, we are also using the same model to discuss the resistive transition broadening in our superconducting samples. The temperature dependence of resistivity in this region is given by Arrhenius equation,

$$\rho(T,B) = \rho_0 \exp[-U_0/k_B T]$$

where, $\rho_0$ is the field independent pre-exponential factor (here normal state resistance at 14 K is taken as $\rho_0$), $k_B$ is the Boltzmann's constant and $U_0$ is TAFF activation energy. $U_0$ can be obtained from the slope of the linear part of an Arrhenius plot in low resistivity region. We have plotted experimental data as $\ln(\rho/\rho_{12})$ vs. $T^{-1}$ in Fig. 6(a) and 6(b) for $PrFe_{0.9}Co_{0.1}AsO$ and $PrFe_{0.85}Co_{0.15}AsO$ samples respectively. The best fitted line to the experimental data gives value of the activation energy ranging from $U_0/k_B$ = 451.7 K to 17.9 K and 1124.34 K to 57.91 K for 10% Co and 15% Co doped samples respectively in the magnetic field range of 0 Tesla - 10 Tesla. Previous investigation of high-$T_c$ superconductors showed that the TAFF activation energy shows different power-law dependences on a magnetic field, i.e. $U_0/k_B \sim H^n$. The magnetic field dependence of activation energy of $PrFe_{0.9}Co_{0.1}AsO$ and $PrFe_{0.85}Co_{0.15}AsO$ are shown in inset of Fig.6 (a) and (b) respectively. The $U_0/k_B$ values for studied samples are an order of magnitude lower than that for high $T_c$ cuprates [27], but comparable to other Fe based pnictide superconductors [28, 29]. This shows that Fe based superconductors are comparatively more robust against magnetic field than the high $T_c$ cuprates.

**Conclusion**

Summarily, it is clear that Co doping at Fe site in $PrFe_{1-x}Co_xAsO$; (a) takes place iso-structurally, (b) fast suppresses both $T^{SDW}(Fe)$ and anomalous $T^N(Pr)$ for low Co content (x < 0.05), (c) introduces robust ($dT_c/dH$ = 0.61K/Tesla, $U_0/k_B$ = 451.7 K to 17.9 K) superconductivity in narrow Co doping level of x between 0.10 to 0.25 and (d) superconductivity disappears for higher Co contents (x > 0.25). Though Pr exhibits anomalous $Pr^{TN}$ of 12K in PrFeAsO due to possible $Pr^{4f}$ and $Fe^{3d}$ interaction, the evolution of superconductivity with Co doping at Fe site i.e., $PrFe_{1-x}Co_xAsO$ is strikingly similar to that as observed for other $REFe_{1-x}Co_xAsO$ (RE = La, Sm, Nd, Gd) compounds.

**Acknowledgement**

Authors acknowledge keen interest and encouragement of their Director Prof. R.C. Budhani. Poonam Rani is supported by CSIR research intern scheme. Anand Pal is financially supported from UGC-Dr. D S Kothari Post Doctoral Fellowship. The work is supported from DAE-SRC outstanding researcher Award scheme to work on search for new superconductors.

**Table 1** Reitveld refined parameters for PrFe$_{1-x}$Co$_x$AsO

| $x$ | $a$ (Å) | $c$ (Å) | Vol. (Å$^3$) | $X^2$ |
|---|---|---|---|---|
| 0 | 3.979(6) | 8.608(9) | 136.35 | 3.22 |
| 0.02 | 3.979(4) | 8.608(7) | 136.33 | 2.31 |
| 0.05 | 3.979(7) | 8.604(8) | 136.28 | 2.52 |
| 0.07 | 3.979(6) | 8.595(6) | 136.13 | 2.72 |
| 0.10 | 3.979(2) | 8.587(3) | 135.97 | 2.38 |
| 0.15 | 3.979(6) | 8.574(5) | 135.79 | 2.40 |
| 0.20 | 3.981(3) | 8.566(5) | 135.78 | 2.06 |
| 0.25 | 3.982(9) | 8.555(8) | 135.66 | 4.17 |
| 0.30 | 3.983(4) | 8.547(6) | 135.63 | 5.95 |
| 0.50 | 3.991(5) | 8.516(2) | 135.68 | 2.07 |
| 0.70 | 4.001(2) | 8.463(5) | 135.44 | 2.79 |
| 1.00 | 4.012(6) | 8.354(2) | 134.51 | 3.40 |

**Figure Captions**

Figure 1:  (a) Room temperature observed and fitted X-ray diffraction patterns for representative samples of PrFe$_{1-x}$Co$_x$AsO (x = 0.0, 0.02, 0.1, 0.5 and 1.0).

(b) Variation of lattice parameter and unit cell volume with Co concentration.

Figure 2:  Resistivity versus temperature ρ(T) plots of representative samples of PrFe$_{1-x}$Co$_x$AsO at zero field. Inset shows the electronic phase diagram of Co-doped PrFe$_{1-x}$Co$_x$AsO.

Figure 3:  (a) Resistivity versus magnetic field ρ(T)H plots for PrFe$_{0.9}$Co$_{0.1}$AsO and (b) PrFe$_{0.85}$Co$_{0.15}$AsO samples respectively.

Figure 4:  (a) The Ginzburge Landau (GL) fitted variation of upper the upper critical field (H$_{c2}$) with temperature for PrFe$_{0.9}$Co$_{0.1}$AsO and (b) PrFe$_{0.85}$Co$_{0.15}$AsO samples respectively at 90%, 50% and 10% criteria.

Figure 5:  (a) Temperature derivative of normalized resistivity for PrFe$_{0.9}$Co$_{0.1}$AsO and (b) PrFe$_{0.85}$Co$_{0.15}$AsO samples respectively.

Figure 6:  (a) Fitted Arrhenius plots of resistivity for PrFe$_{0.9}$Co$_{0.1}$AsO and (b) PrFe$_{0.85}$Co$_{0.15}$AsO samples respectively. U$_0$ dependence of magnetic field is shown in their respective insets.

Figure 1(a)

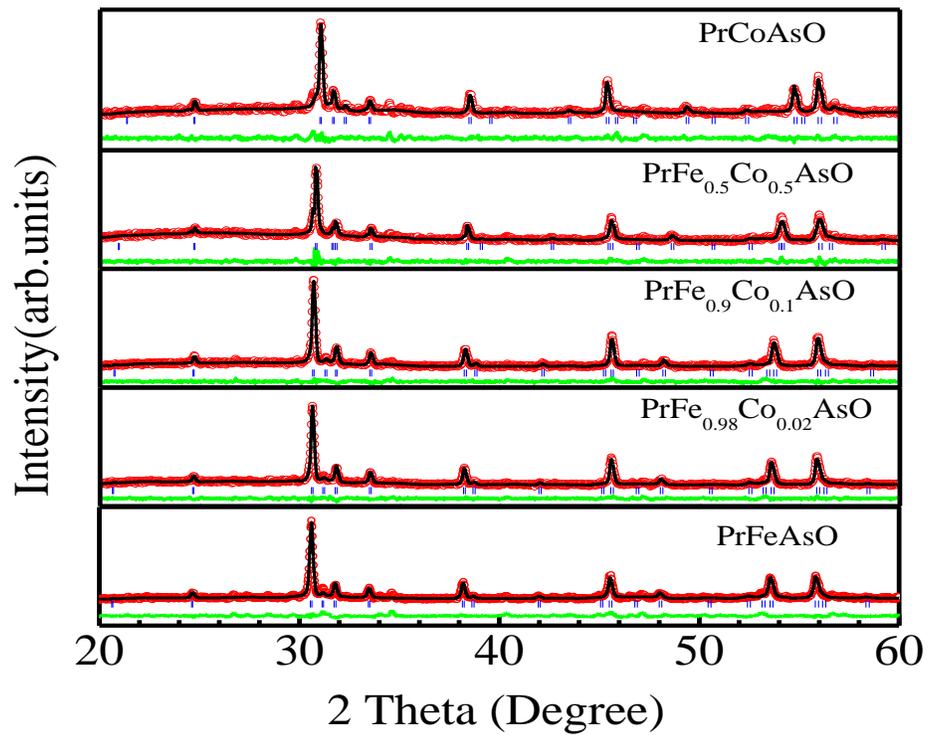

Figure 1(b)

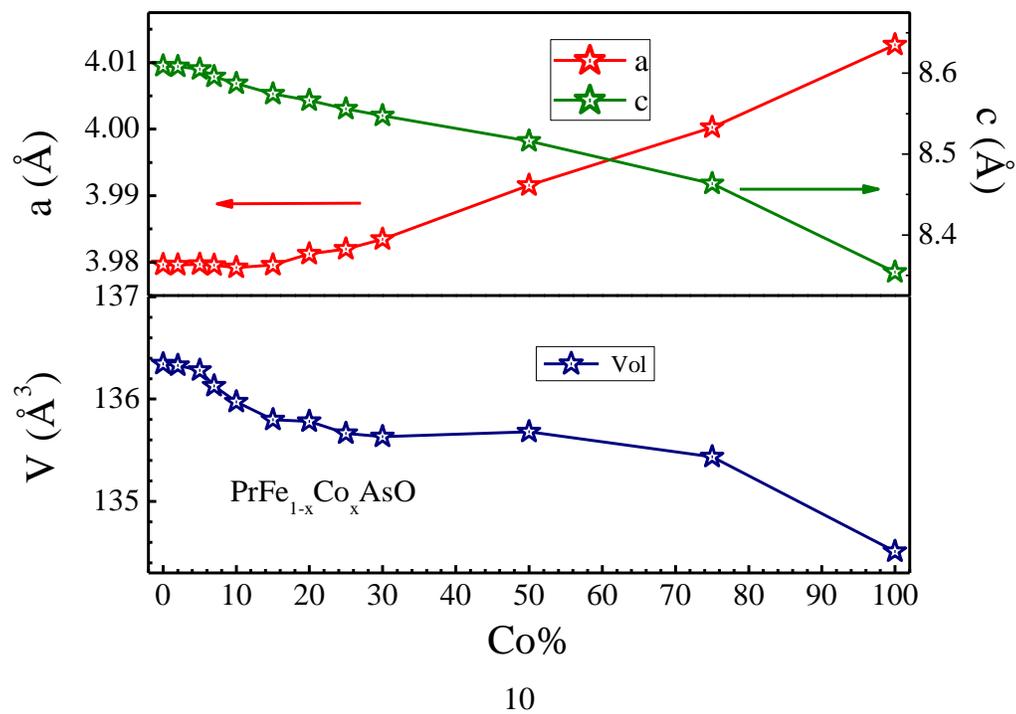

Figure 2

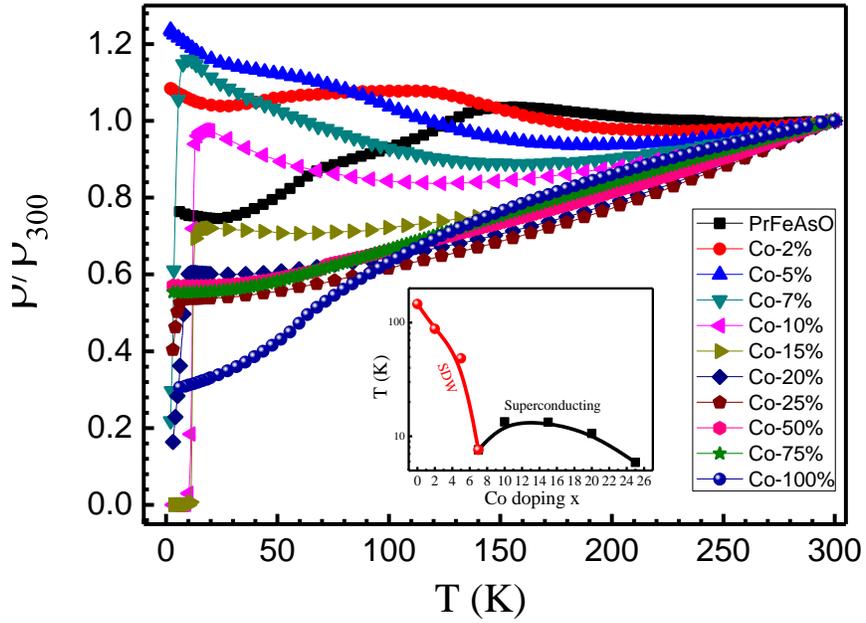

Figure 3 (a)

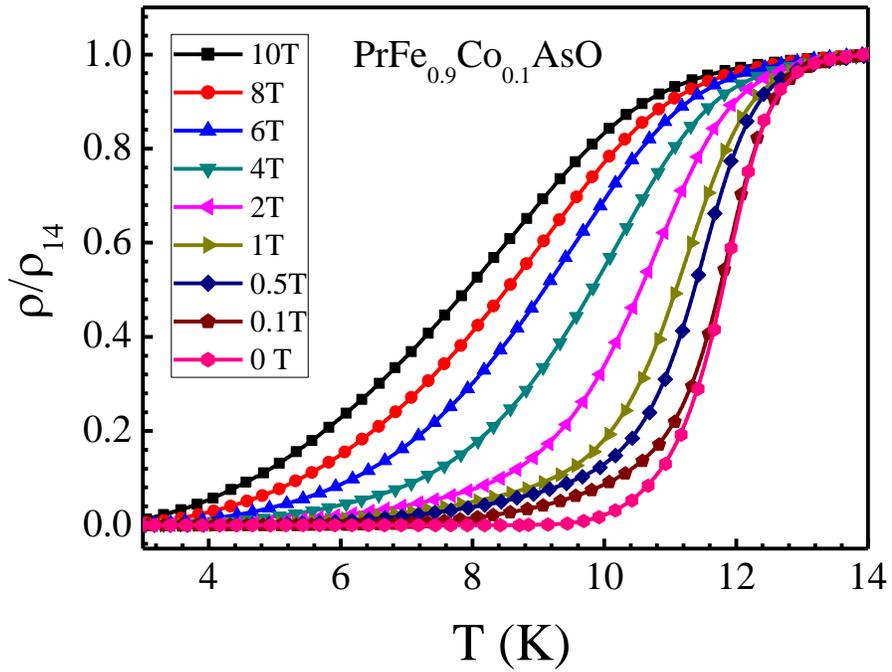

Figure 3 (b)

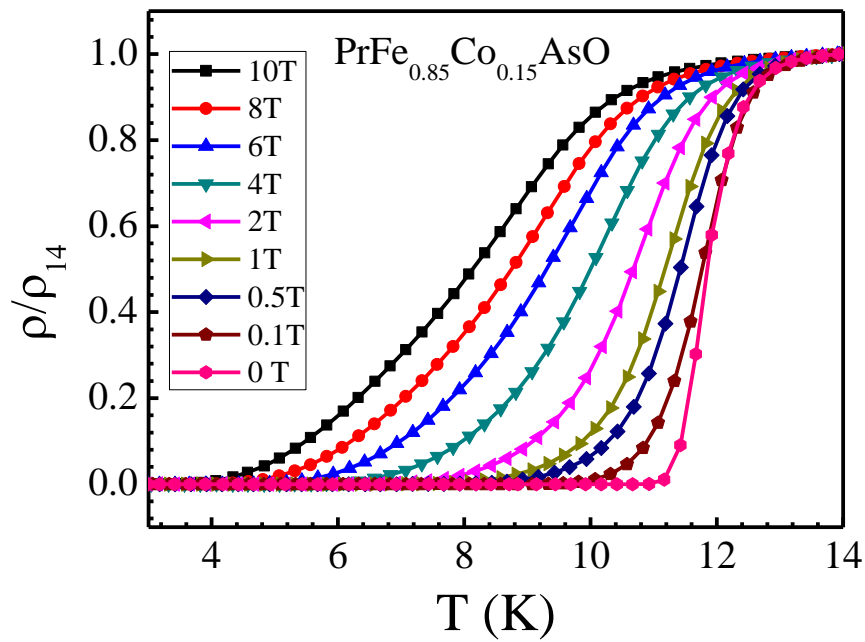

Figure 4 (a)

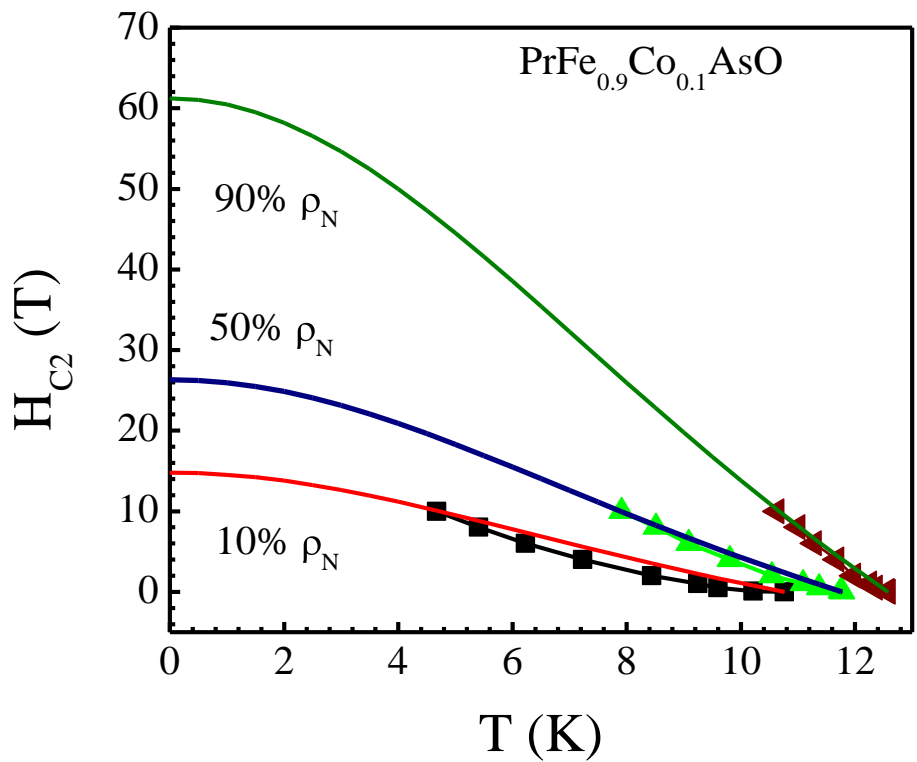

Figure 4 (b)

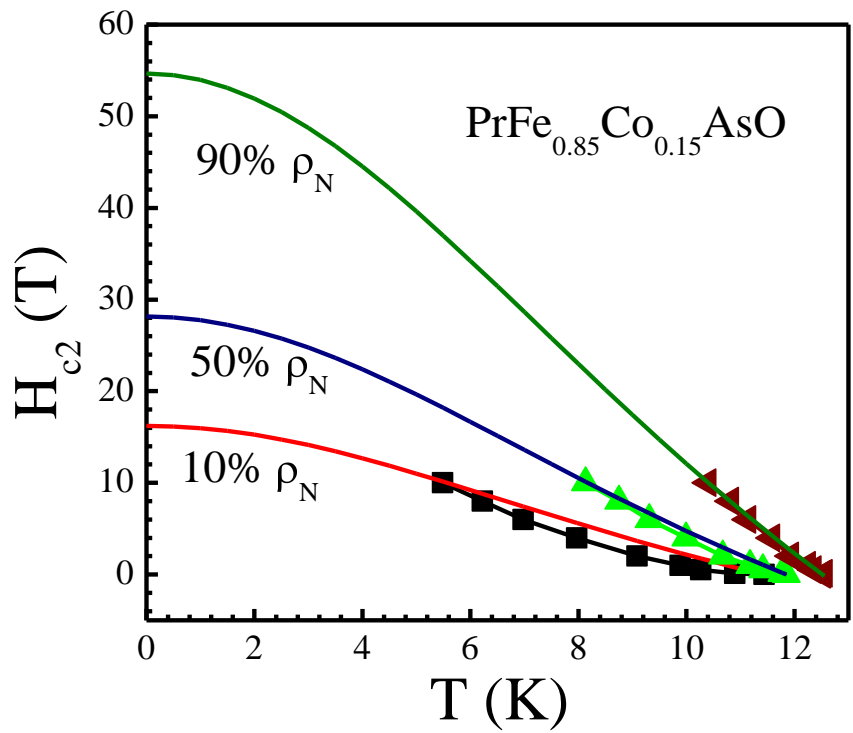

Figure 5 (a)

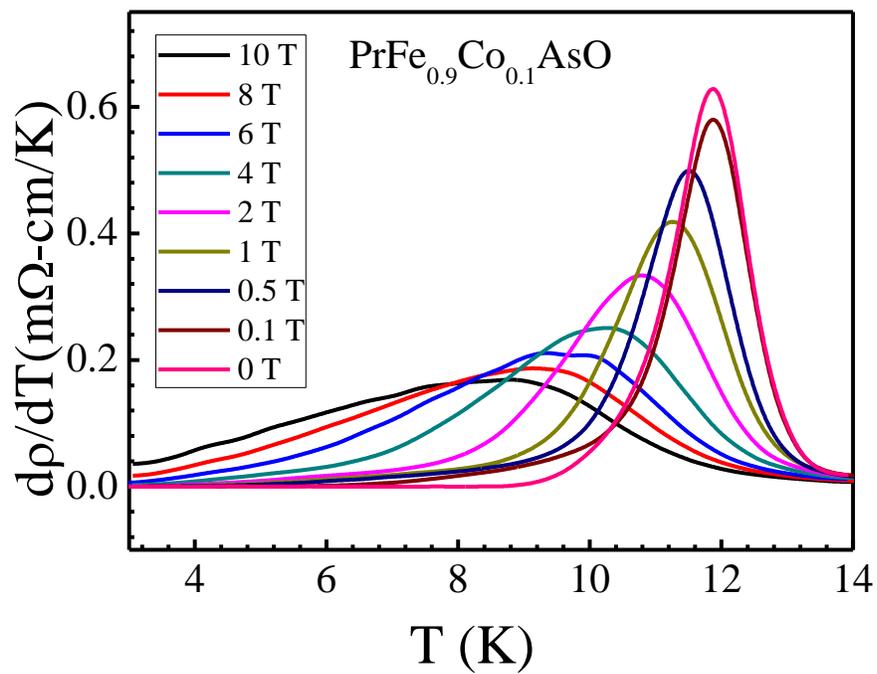

Figure 5 (b)

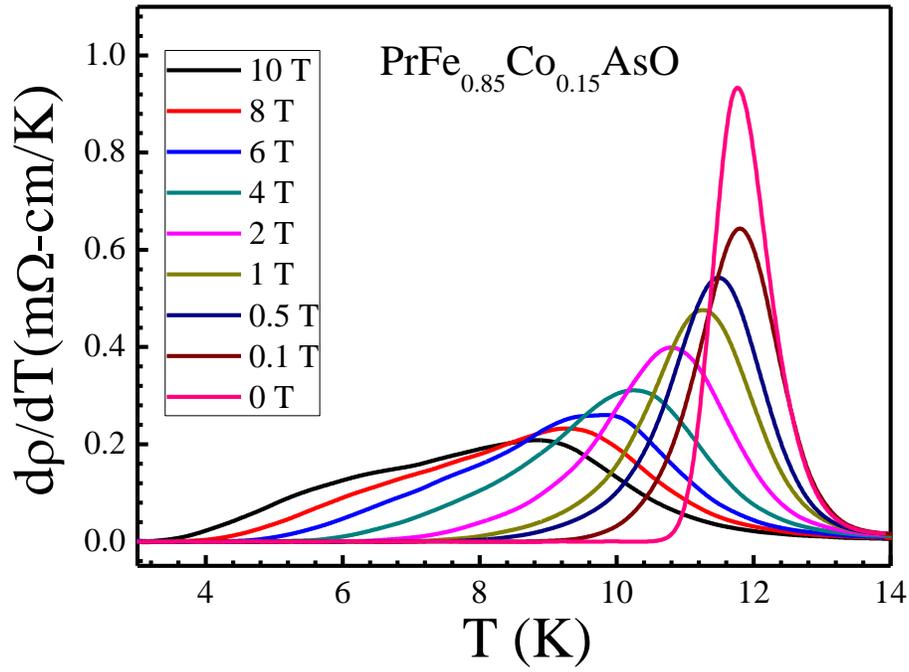

Figure 6 (a)

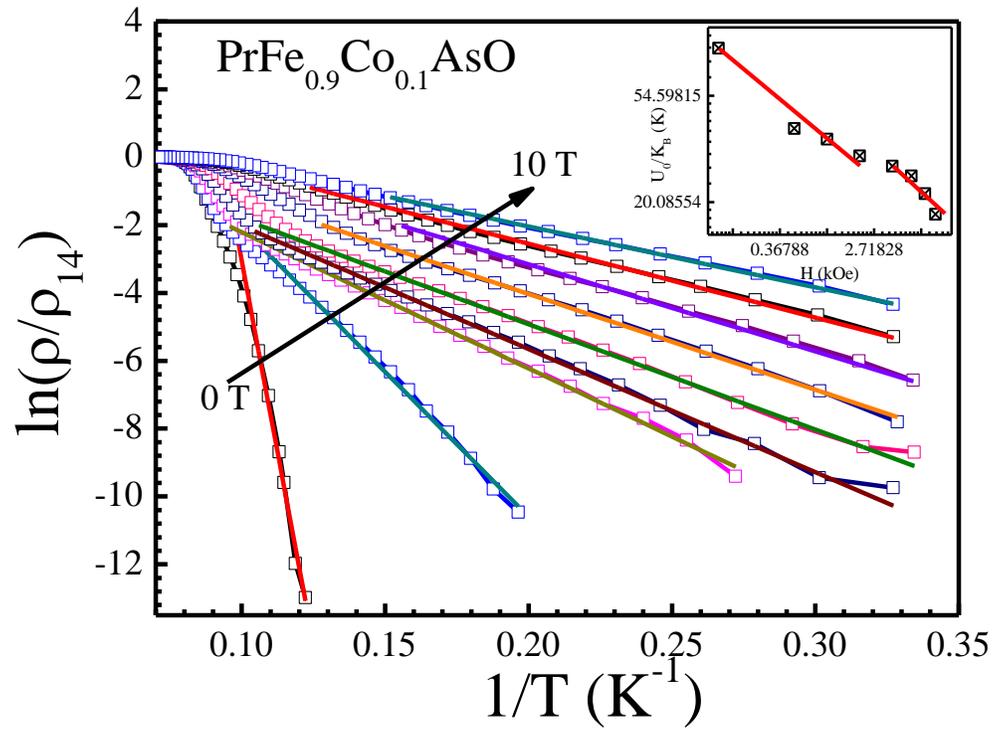

Figure 6 (b)

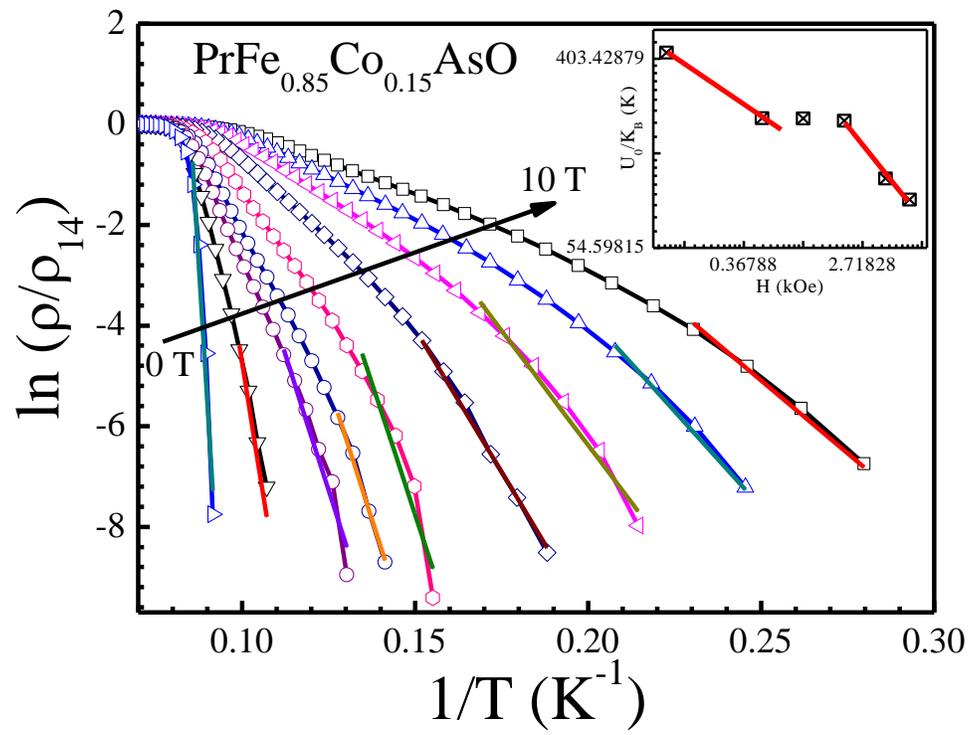